\documentclass[%
 reprint,
 superscriptaddress,
 showpacs,
 amsmath,amssymb,
 aps,
 prl,
]{revtex4-1}

\usepackage{graphicx}
\usepackage{dcolumn}
\usepackage{bm}

\usepackage[utf8]{inputenc}
\usepackage[T1]{fontenc}
\usepackage{mathptmx}

\usepackage{siunitx}
\usepackage{color}

\DeclareMathOperator{\sinc}{sinc}

\begin{document}

\preprint{AIP/123-QED}

\title[]{Temperature insensitive type II quasi-phasematched spontaneous parametric downconversion}

\author{Pan Xin-Yi}
 \affiliation{Centre for Quantum Technologies, 3 Science Drive 2, National University of Singapore, 117543 Singapore}
 
\author{Christian Kurtsiefer}
\affiliation{Centre for Quantum Technologies, 3 Science Drive 2, National University of Singapore, 117543 Singapore}
\affiliation{Department of Physics, National University of Singapore, Blk S12, 2 Science Drive 3, 117551 Singapore}

\author{Alexander Ling}
 \email{alexander.ling@nus.edu.sg}
\affiliation{Centre for Quantum Technologies, 3 Science Drive 2, National University of Singapore, 117543 Singapore}
\affiliation{Department of Physics, National University of Singapore, Blk S12, 2 Science Drive 3, 117551 Singapore}

\author{James A. Grieve}
 \email{james.grieve@tii.ae}
\affiliation{Centre for Quantum Technologies, 3 Science Drive 2, National University of Singapore, 117543 Singapore}
\affiliation{Quantum Research Centre, Technology Innovation Institute, Abu Dhabi, UAE}

\date{\today}

\begin{abstract}
The temperature dependence of the refractive indices of potassium titanyl phosphate (KTP) are shown to enable quasi-phasematched type II spontaneous parametric downconversion (SPDC) with low temperature sensitivity. Calculations show the effect to be maximised for emission of photons at around \SI{1165}{\nano\meter}, as well as producing potentially useful regions for wavelengths throughout the telecommunications bands. We demonstrate the effect experimentally, observing temperature-insensitive degenerate emission at \SI{1326}{\nano\meter}, within the telecommunications O band. This result has practical applications in the development of entangled photon sources for resource-constrained environments, and we demonstrate a simple polarization entangled source as a proof of concept.
\end{abstract}

\maketitle



\noindent Entangled photon sources form a core building block of a wide variety of quantum communication protocols~\cite{Ekert1991,Briegel1998,Wehner2018,Ribordy2000}. For practical applications involving optical fiber, sources operating in the low-attenuation telecommunications bands are highly desirable~\cite{Tanzilli2001}, with a number of mature and even commercial examples operating in the so-called C-band (centred at \SI{1550}{\nano\meter}) where transmission losses are minimized. For short-haul distances and metropolitan networks, operation can also be viable in the O-band (centred at \SI{1310}{\nano\meter}), where a relative increase in attenuation can trade off against the prospect of multiplexing quantum signals with classical traffic in the C-band~\cite{Shi2020,Xiang2019}.

The workhorse technique for generating entangled photons is based on Spontaneous Parametric Down Conversion (SPDC) in bulk non-linear optical materials~\cite{Burnham1970,Klyshko1970,Kwiat1995,Anwar2020}. In addition to critically phase-matched systems such as Beta Barium Borate~\cite{Kwiat1995} (BBO), many sources make use of quasi phase-matched processes in periodically poled crystals~\cite{Hum2007}, for example Lithium Niobate~\cite{Tanzilli2001} (PPLN) or Potassium Titanyl Phosphate~\cite{Fiorentino2007} (PPKTP). In order to maintain the desired phasematching characteristics these poled materials typically need to be temperature regulated to within \SI{0.1}{\celsius}.

While this is readily accomplished under laboratory conditions, it may be desirable to relax these constraints for field deployment, to improve device stability or reduce mass or power requirements~\cite{Chandrasekara2015}. In this paper, we show by simulation how the thermo-optic properties of KTP give rise to reduced temperature sensitivity for quasi-phasematched downconversion into a range of wavelengths overlapping the popular telecommunications bands. A source of O-band polarization entangled photons is also realized experimentally, in which the photon pairs are generated in a SPDC process stable over an \SI{80}{\celsius} range.


\begin{figure}[b!]
\centering
\includegraphics[width=\linewidth]{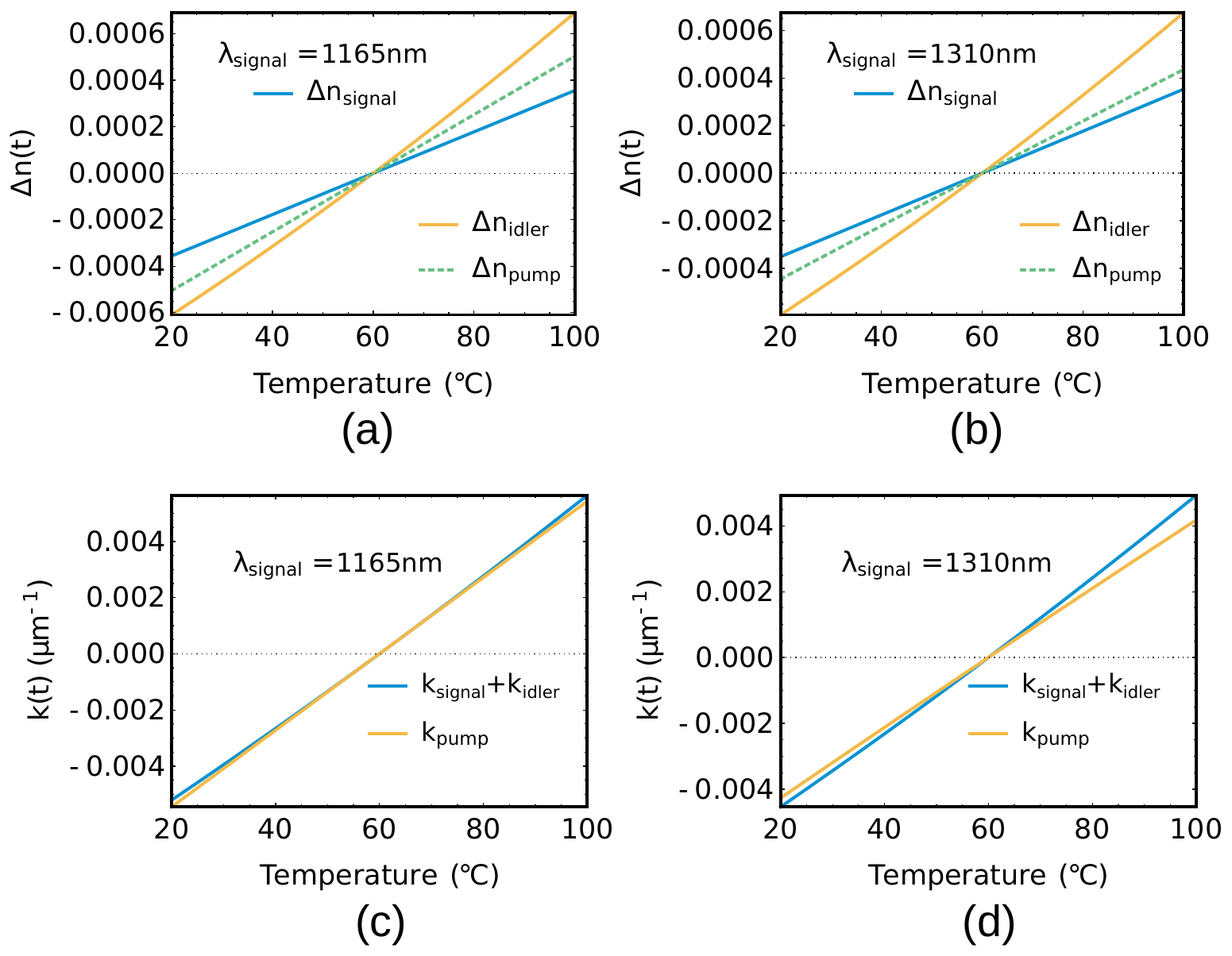}
\caption{(a,b) Calculated thermally induced changes in the refractive index for pump, signal and idler photons in a Type II quasi-phasematched SPDC process in x-cut PPKTP, using published thermo-optic coefficients for $Y$ and $Z$ axes from~\cite{Kato2002,Emanueli2003}. Curves are plotted taking \SI{60}{\celsius} as a reference, for emission at (a) \SI{1165}{\nano\meter}, poling period \SI{102.2}{\micro\meter} and (b) \SI{1310}{\nano\meter}, poling period \SI{54.6}{\nano\meter}. Poling periods calculated to produce degenerate emission at \SI{60}{\celsius}. Corresponding changes in wavevector ($k(t)$) for pump and (summed) signal and idler photons are shown in (c,d).}
\label{fig: index_and_k_vs_t}
\end{figure}

\vspace{0.3cm}
\noindent Spontaneous parametric downconversion (SPDC) is a process in which photons can split into pairs of lower energy ``daughter'' photons, obeying conservation of momentum and energy laws. These constraints define the spectral properties of the generated fields. In particular, conservation of momentum within the nonlinear crystal gives rise to phasematching conditions, in which the generated wavelengths depend strongly upon the refractive indices experienced by the downconverted photons~\cite{Christ2013}. Where the phasematching conditions cannot be satisfied directly by the properties of the medium, it may be possible to compensate the unbalanced phase by engineering a periodic inversion of the crystal symmetry. This technique is known as quasi-phasematching, with the relevant phase relationship given by

\begin{equation}
\label{eqn:kvectors}
    \Delta k = k_{pump} - k_{signal} - k_{idler} - \frac{2 \pi}{\Lambda},
\end{equation}

\noindent where $k=2\pi n/\lambda$ is the wavevector in the medium, and $\Lambda$ is the poling period: the domain length of the periodic inversion~\cite{Fejer1992}. The process is considered perfectly phase-matched when the imbalance ($\Delta k$) is equal to zero. The intensity of the downconverted fields in a crystal of length $L$ is then given by

\begin{equation}
\label{eqn:sinc}
    I_{SPDC} = \sinc^2(\Delta k \, L /2).
\end{equation}

The spectral characteristics of the SPDC photons are in general highly temperature sensitive, due to the temperature sensitivity of the refractive index and to a lesser extent the impact of thermal expansion upon the poling period.

In Type II collinear processes in PPKTP, serendipitous correlations between the thermo-optic effect in the Y and Z refractive indices for certain values of $\lambda_{pump,signal,idler}$ can give rise to low sensitivity to temperature. While this has been previously noted for critical phasematched second harmonic generation~\cite{Kato92}, these correlations may also be leveraged in quasi-phasematched processes, greatly extending their utility. Our calculations are based on the KTP characterizations performed by \emph{Kato and Takaoka}~\cite{Kato2002} and \emph{Emanuelli and Ady}~\cite{Emanueli2003}, with the latter including a model for the thermal dependence of the refractive index. The effect appears to be most dramatic for degenerate emission at \SI{1165}{\nano\meter}, where it is almost invariant over a \SI{100}{\celsius} region. It is also visible in operationally useful wavelengths spanning the Telecom O and C bands.

The calculated temperature dependence of the refractive index experienced by pump, signal and idler fields in a Type II collinear SPDC process in an x-cut KTP crystal are plotted in Figure~\ref{fig: index_and_k_vs_t}(a) and (b) for degenerate emission at \SI{1165}{\nano\meter} and \SI{1310}{\nano\meter} respectively, taking \SI{60}{\celsius} as a reference. It is clear that variation in the $Y$ index at the pump wavelength lies between the variation of the $Y$ and $Z$ indices at the signal/idler wavelengths, leading to a degree of compensatory behaviour in the resulting wavevectors. This is shown for the same wavelengths in Figure~\ref{fig: index_and_k_vs_t}(c,d), where the sum of changes in the signal and idler wavevectors approximately matches that of the change in pump wavevector over a wide temperature range.

\begin{figure}[t]
\includegraphics[width=\linewidth]{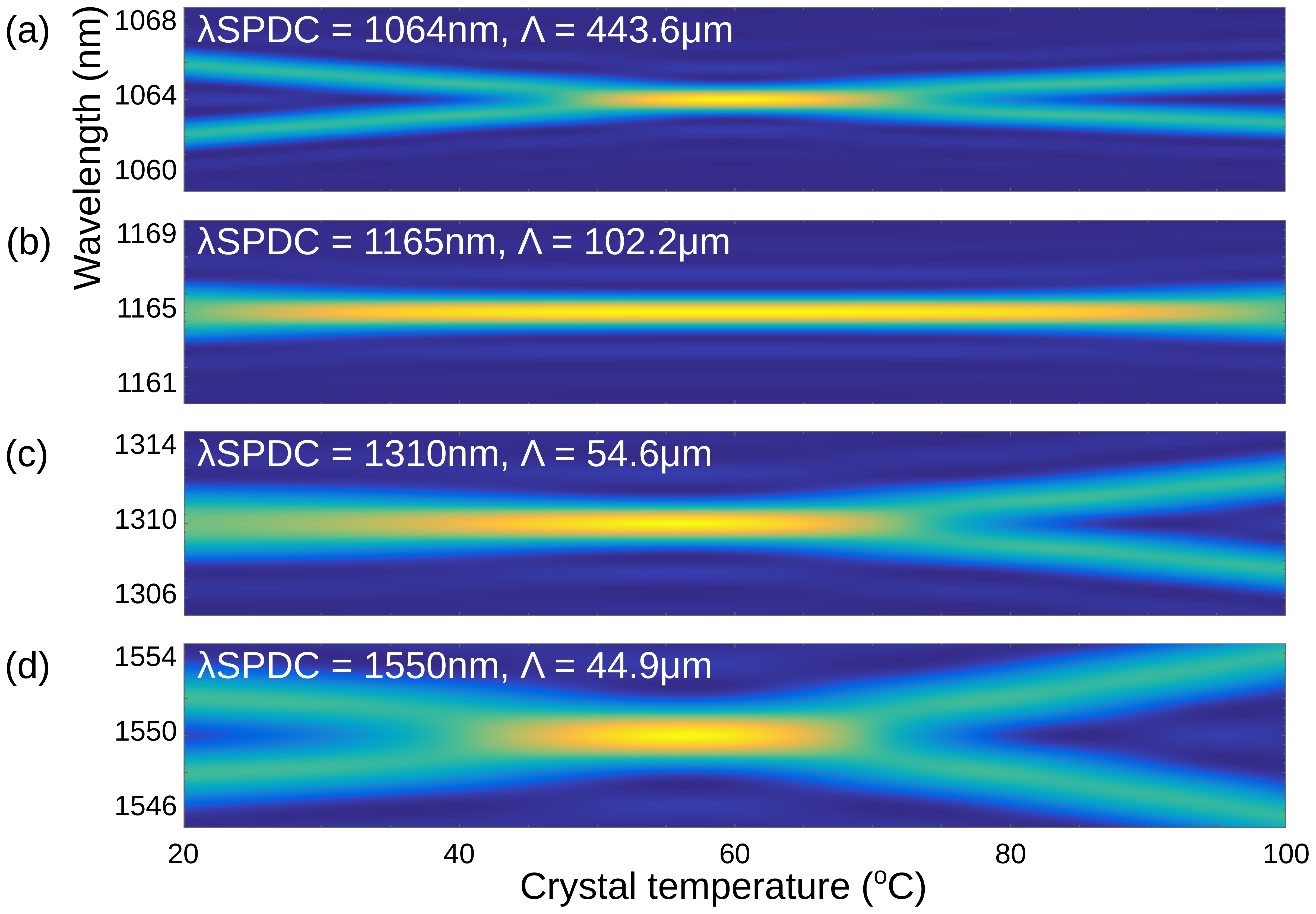}
\caption{Calculated SPDC spectra as a function of crystal temperature for a variety of commonly used wavelengths. The spectrum is simulated for \SI{10}{\milli\meter} PPKTP crystals with poling periods ($\Lambda$) calculated to produce degenerate Type II SPDC at \SI{60}{\celsius} (see plot labels). We observe near-degenerate regions in which signal and idler photon spectra overlap for a wide range of temperatures. While the effect is most pronounced for downconversion to \SI{1165}{\nano\meter} (b), it is seen in all calculations with target wavelengths above $\sim$\,\SI{1000}{\nano\meter}.}
\label{fig: heatmap_crystal_sim}
\end{figure}

Quasi-phasematched, Type II SPDC spectra are calculated for a range of interesting wavelengths, with poling periods calculated to produce degenerate collinear emission at \SI{60}{\celsius} and the effects of thermal expansion included~\cite{Emanueli2003}. The results are plotted in Figure~\ref{fig: heatmap_crystal_sim}, and include \SI{1064}{\nano\meter} (a), \SI{1165}{\nano\meter} (b), \SI{1310}{\nano\meter} (c) and \SI{1550}{\nano\meter} (d). While the temperature insensitive behaviour is most dramatically observed at \SI{1165}{\nano\meter}, the wide plateau also observed at \SI{1310}{\nano\meter} may be of more practical interest as it lies at the centre of the telecommunications O band.



\vspace{0.1cm}

\noindent In order to observe temperature-insensitive emission, we select a PPKTP crystal with \SI{54.05}{\micro\meter} poling period which is specified by the manufacturer for Type II second harmonic generation at \SI{1310}{\nano\meter}. While this is similar to the \SI{54.6}{\micro\meter} predicted by our calculations (see Figure~\ref{fig: heatmap_crystal_sim}(c)), small discrepancies of this sort are commonly attributed to the manufacturer using proprietary dispersion relationships measured for their own materials, which may differ slightly from those available in the literature.

\begin{figure}[b]
\includegraphics[width=\linewidth]{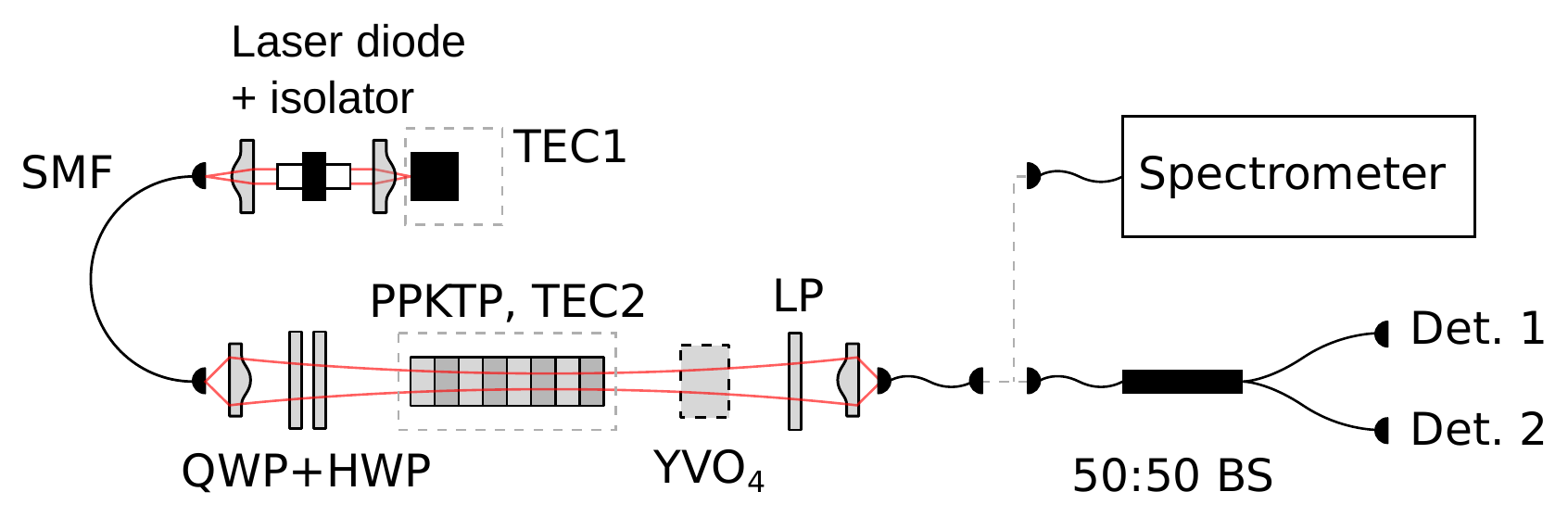}
\caption{Experimental schematic. We use a single longitudinal mode laser diode to pump a PPKTP crystal, with downconverted photons collected into a single mode fiber (SMF28, Corning) using appropriate focusing and collection lenses. Residual pump light is removed prior to the collection optics by a long-pass filter (LP). In order to investigate temperature-insensitive quasi-phasematching, the laser diode and PPKTP crystal are mounted on independent temperature controlled stages (TEC1, TEC2). The output fiber may be routed to a spectrometer, or to a fiber beamsplitter and avalanche photodiodes. When operated as a polarization entangled photon source, a single YVO$_4$ crystal is placed in the path of the downconverted beam to compensate for signal and idler phase differences.}
\label{fig: schematic}
\end{figure}

We assemble a correlated photon pair source following the schematic in Figure~\ref{fig: schematic}. The PPKTP crystal is pumped from a fiber-coupled laser diode, with downconverted photons emitted in a collinear direction collected into a single mode fiber using appropriate collection optics~\cite{Dixon2014}. Residual pump light is removed using a long pass filter.

When operating the system with all components at room temperature, the source emits non-degenerate photon pairs centred at \SI{1316}{\nano\meter} (pump wavelength \SI{658}{\nano\meter}). While most quasi-phasematched sources in the literature employ temperature tuning of the downconversion crystal to achieve degenerate emission, this is obviously not practical in our system (see Figure~\ref{fig: heatmap_crystal_sim}). Instead, we tune the SPDC spectrum via the pump wavelength. This is achieved by varying the temperature of our pump laser, a free running laser diode designed to produce a single longitudinal mode (HL6501MG, Hitachi).

\begin{figure}[h]
\includegraphics[width=\linewidth]{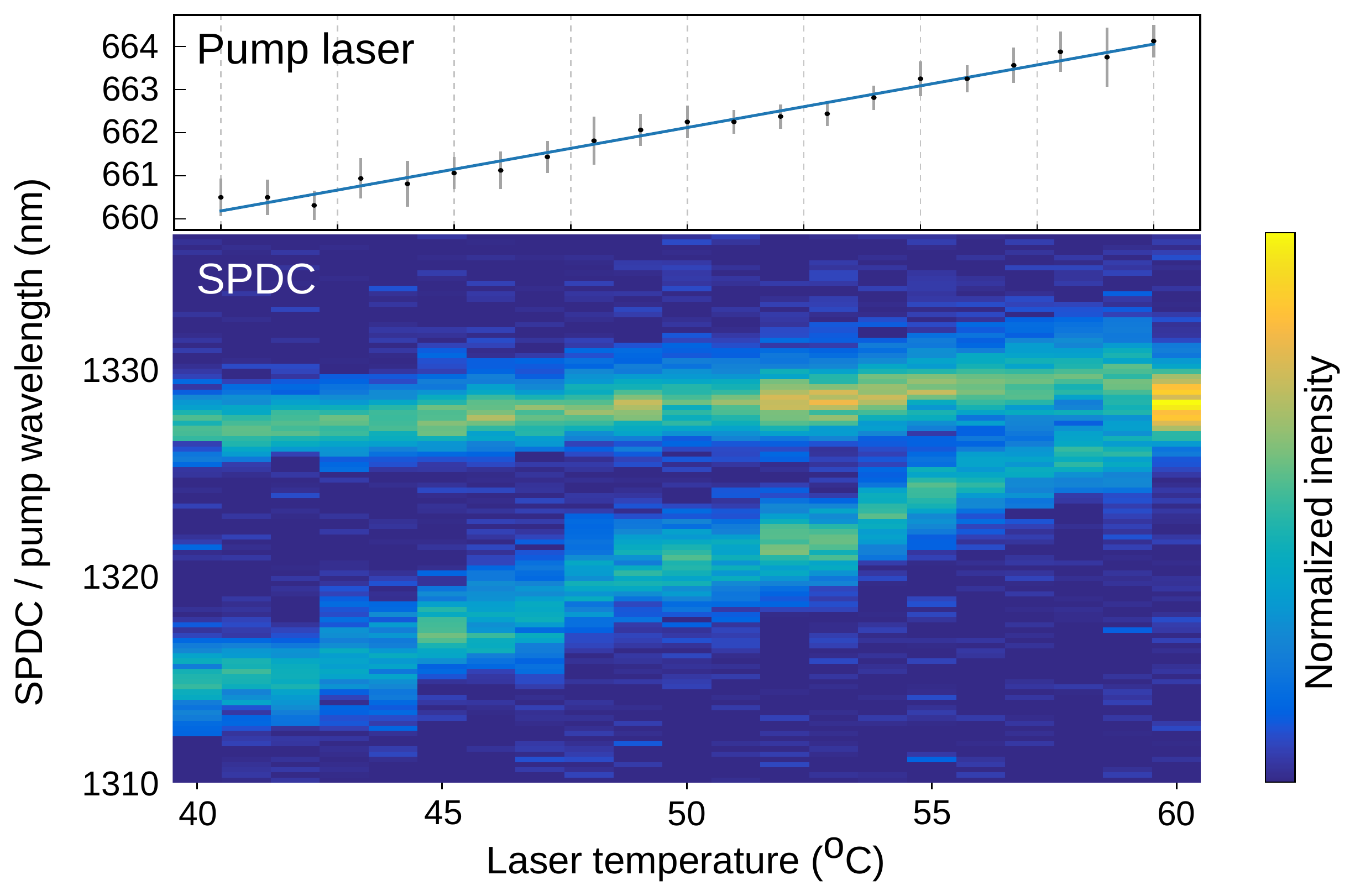}
\caption{Measured SPDC spectrum as a function of pump laser temperature. The spectrum is obtained at a crystal temperature of \SI{22.0}{\celsius}, at input power \SI{7.3(2)}{\mW}. We observe the convergence of signal and idler wavelength with increasing pump laser temperature (wavelength), with a degenerate condition reached at a laser temperature of approximately \SI{60.2}{\celsius}. The upper panel shows the pump laser wavelength extrapolated from the SPDC data, exhibiting a linear relationship between laser temperature and wavelength, with a trend of \SI{0.18}{\nano\meter\per\celsius}.}
\label{fig: heatmap_laser}
\end{figure}

The observed spectrum of downconverted photons as a function of pump laser temperature is plotted in Figure~\ref{fig: heatmap_laser}. Our results are consistent with a continuously varying pump wavelength, and we infer a tuning rate of \SI{0.18}{\nm\per\celsius}. The production of degenerate photons is observed at a laser temperature of approximately \SI{60.2}{\celsius}, corresponding to a pump laser wavelength of \SI{664}{\nano\meter}.

\begin{figure}[h]
\includegraphics[width=\linewidth]{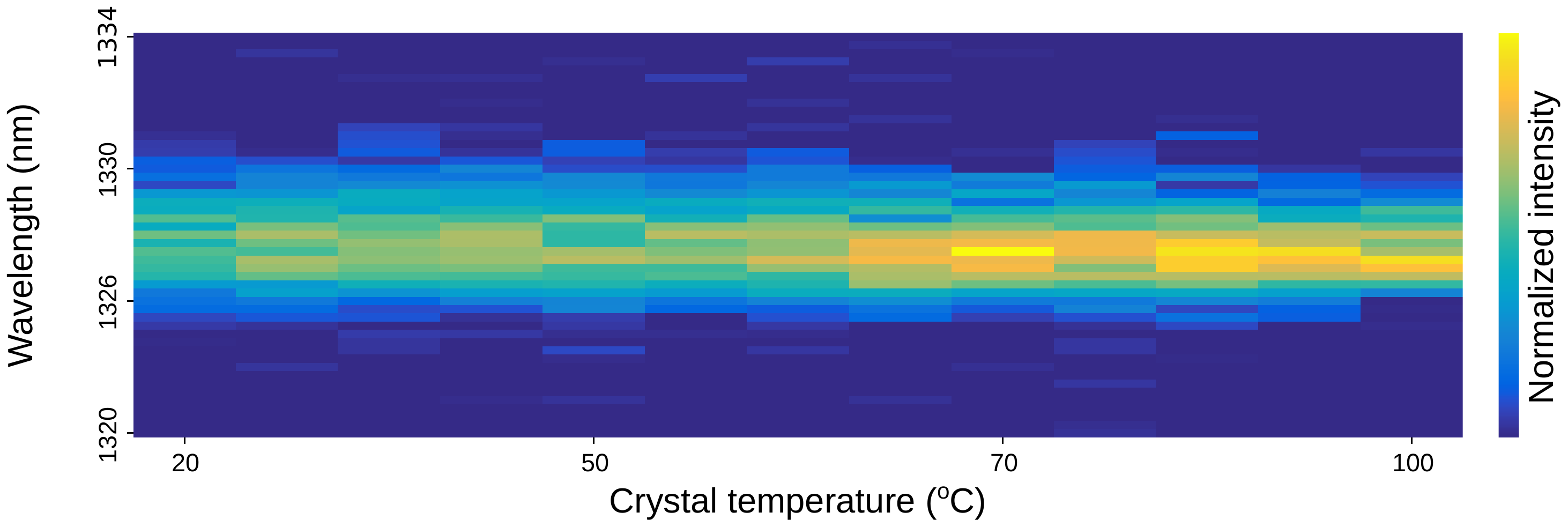}
\caption{Measured SPDC Spectrum as a function of crystal temperature. The spectrum is obtained at a constant pump laser temperature of \SI{59.6}{\celsius}, for temperatures in the range \SI{20.0}{\celsius} to \SI{110.0}{\celsius} We observe relatively little sensitivity to this parameter, with signal and idler photons indistinguishable across this range.}
\label{fig: heatmap_crystal}
\end{figure}

To confirm the relative insensitivity of our source to the PPKTP crystal temperature, we measure the SPDC spectrum as the crystal is heated over a range of \SIrange{20}{110}{\celsius}. The results are shown in Figure~\ref{fig: heatmap_crystal}, and confirm a low variation, with the signal and idler photons spectrally indistinguishable across this interval. Limitations in our experimental apparatus prevented us from increasing the temperature further.

\noindent Type II degenerate SPDC systems are natural candidates for the construction of polarization entangled photon sources. After collection into a single mode fiber, downconverted photon pairs may be split into two fibers in a probabilistic manner. Since the Type II SPDC process produces orthogonally polarized photons, interference between the two possible polarization/mode pairings results in the creation of the state

\begin{equation}
    |\Psi^+\rangle = \frac{1}{\sqrt{2}} (|H_1 V_2\rangle + |V_1 H_2\rangle),
\end{equation}

\noindent where $|H_{1}\rangle$ denotes a horizontally polarized photon in fiber 1, and $|V_2\rangle$ a vertically polarized photon in fiber 2, and vice versa. This mechanism is similar to that described in~\cite{Shih94,Chen18}, in that both components of the state arise from the same SPDC process. Consequently, there is no phase introduced between the $|HV\rangle$ and $|VH\rangle$ components (though some may later be introduced within the optical fibers).

The birefringent, dispersive nature of the PPKTP crystal causes each photon pair travelling through the crystal to incur a relative phase. Unless compensated, this phase difference will result in the creation of a mixed state over the downconverted photons' spectrum, effectively obscuring the polarization entanglement. We make use of a second birefringent crystal (Yttrium orthovanadate, YVO$_4$) to compensate for this phase difference, placing it in the beam after the PPKTP crystal with its extraordinary axis oriented perpendicular to that of the PPKTP (for more detailed discussion of ``longitudinal phase'' and this compensation technique, see for example references~\cite{Altepeter2005,Rangarajan2009}).

\begin{figure}[h]
\includegraphics[width=\linewidth]{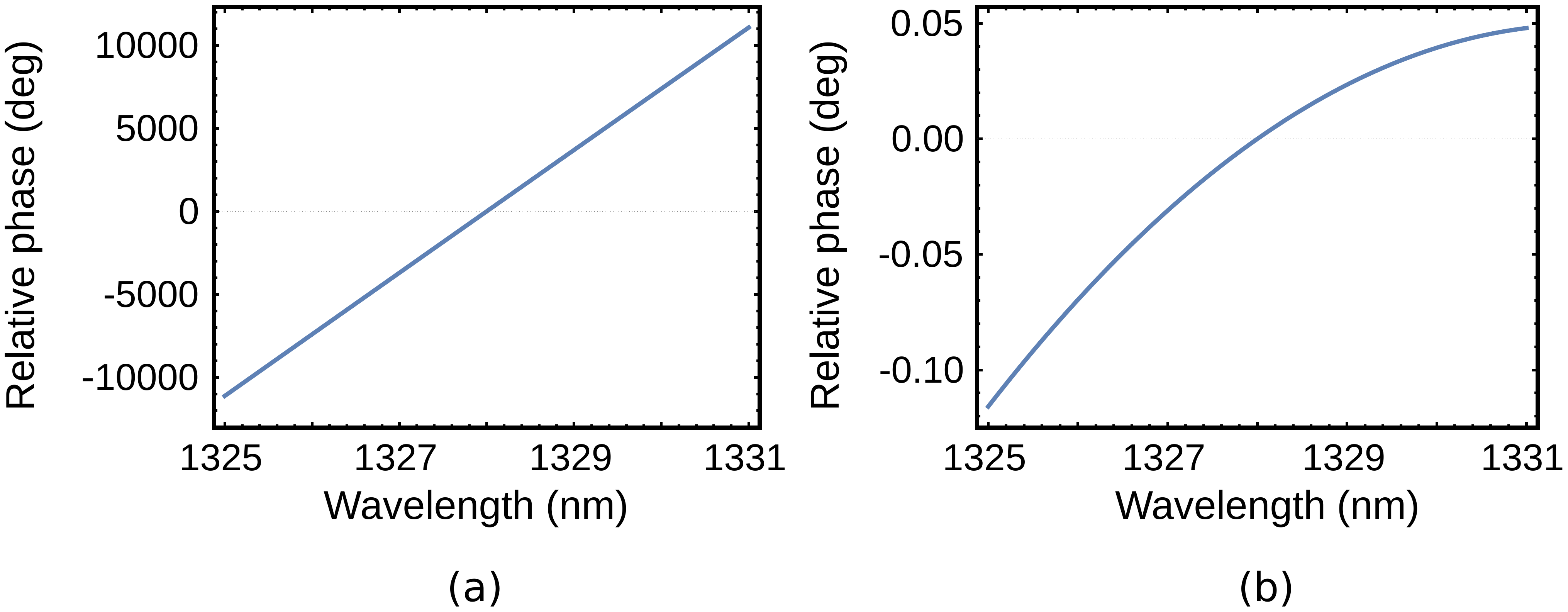}
\caption{Calculated phase differences between signal and idler photons as a function of wavelength, for both uncompensated (a) and compensated (b) scenarios. Calculations were performed for orthogonally polarized photons produced at the centre of a \SI{10}{\milli\meter} PPKTP crystal. For the compensated phase illustrated in (b), photons are further propagated through a \SI{4.34}{\milli\meter} YVO$_4$ crystal, oriented with its extraordinary axis perpendicular to that of the PPKTP.}
\label{fig: phase_calcs}
\end{figure}

To estimate the acquired phase, we take the centre of the PPKTP crystal to be the point at which most of the collected SPDC photon pairs originate (pairs produced before or after this point will be under- or over-compensated accordingly). The phase difference between signal and idler photons as a function of their wavelength is shown in Figure~\ref{fig: phase_calcs}(a), assuming co-propgatation through \SI{5}{\milli\meter} of PPKTP and taking the degenerate wavelength of \SI{1328}{\nano\meter} as a zero reference. Solving for the minimum phase difference, we obtain an optimum YVO$_4$ length of \SI{4.34}{\milli\meter}, with the resulting compensated phase reduced below $\pm$0\SI{.06}{\degree} over the SPDC spectrum as shown in Figure~\ref{fig: phase_calcs}(b).

Our photon pair source is again assembled as described in Figure~\ref{fig: schematic}, with the addition of the YVO$_4$ crystal for phase compensation. Generated photon pairs are split probabilistically using a fused fiber beamsplitter, and detected using InGaAs avalanche photodiodes. Events are recorded using a home-built timestamping system with \SI{0.125}{\nano\second} resolution, and we configure the system to identify coincident detection events within a \SI{8}{\nano\second} window, effectively post-selecting only the outcomes in which the photons are found in separate fibers.

\begin{figure}[b]
\vspace{0.3cm}
\includegraphics[width=\linewidth]{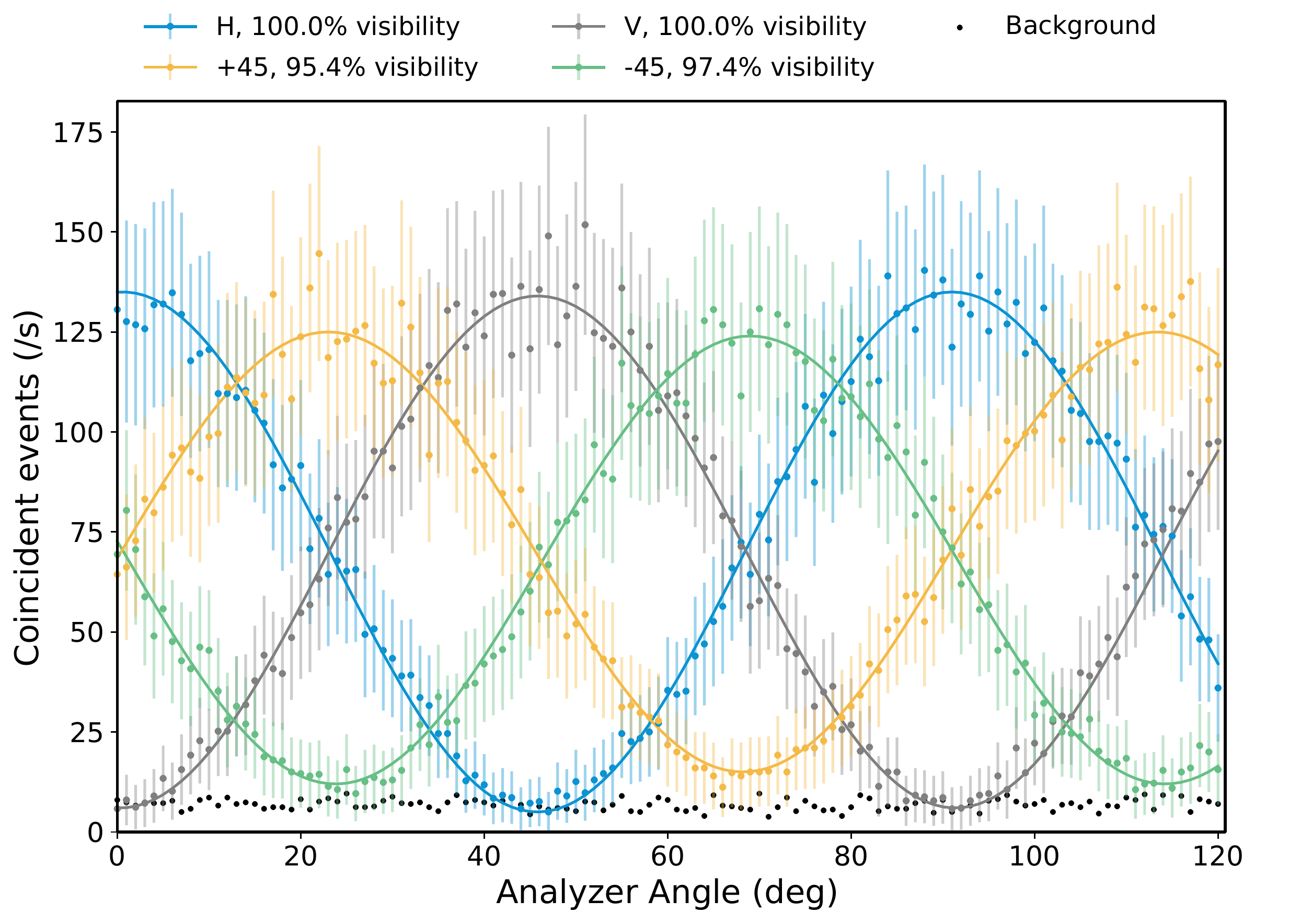}
\caption{Characterization of source polarization enatanglement. We plot the coincident event rate obtained as a function of analyzer angle (signal arm) with the idler polarizer set to H, V, +45 and -45 states. As the data is obtained using a timestamp-based cross-correlation system with 8ns coincidence window, we are able to make an independent assessment of the background (or "accidental") event rate at each point.}
\label{fig: visibility_4bases}
\end{figure}

In order to evaluate the polarization entanglement of the source, we route the two output fibers to a polarization tomography system, with fiber birefringence mitigated using manual polarization controllers. By fixing the idler polarizer to the H, V, +45 and -45 bases and scanning the signal polarizer, we are able to reconstruct the polarization correlations as shown in Figure~\ref{fig: visibility_4bases}. Our timestamp-based signal processing electronics enable the direct measurement of background (or ``accidental'') correlations, and using this data we infer a visibility of approximately 100.0\% in the H and V (production) bases. The visibility in the +45 and -45 bases is 95.4\% and 97.4\% respectively, giving us a fidelity of 98.2\% to the $\Psi^+$ state.

In this proof-of-concept source, the overall pairwise brightness was relatively low as the photon pairs were separated by chance using a fused 50:50 beamsplitter. For a performance improvement, the separation could be conducted instead using momentum correlations~\cite{Perumangatt2020}. One aspect of this geometry is that the pump laser with the correct wavelength must be used. We have achieved this by temperature tuning the laser and then maintaining its temperature stability to within \SI{0.5}{\celsius}. In the future, power requirements could be further optimized by using optical feedback methods (e.g. with a grating~\cite{Ricci1995}) so that the laser could maintain the correct wavelength passively.

While our work has concentrated on degenerate emission, the same principles should also hold for non-degenerate SPDC. For these configurations, efficient separation of signal and idler photons can be achieved by using appropriate dichroic mirrors. It may also be possible to harness the reduced sensitivity to temperature when operating in a non-collinear collection scheme, where an alternative approach to temperature stability has been demonstrated previously in a Type 0 SPDC experiment~\cite{Jabir2017}. In contrast to this work, our technique does not show a significant influence of crystal temperature upon source brightness.



In conclusion, we have described and demonstrated a regime in which quasi-phasematched SPDC can be utilized reliably without stringent temperature control, and still produce high quality entangled photon pairs. We focused our experimental efforts primarily on a source of photon pairs in the telecommunication O-band. However, a region of relaxed temperature requirements is also predicted in the commonly used C-band (see Figure~\ref{fig: heatmap_crystal_sim}(d)) and to a much lesser extent the L-band. This finding can be used to increase the utility and reliability of entangled photon systems for field deployment, and opens up a new class of entangled photon sources for applications.

\section*{Acknowledgments}

This research was carried out at the Centre for Quantum Technologies, National University of Singapore, and supported by the National Research Foundation, Prime Minister’s Office, Singapore under its Corporate Laboratory@University Scheme, National University of Singapore, and Singapore Telecommunications Ltd.

\bibliography{bibliography}

\end{document}